\documentclass[preprint,prl,10pt,twocolumn]{revtex4}%
\usepackage{amsmath}
\usepackage{graphicx}
\usepackage{amsfonts}
\usepackage{amssymb}%
\begin{document}
%
\preprint{Marc Durand}%
%

\title
[Optimizing the bulk modulus]{Optimizing the bulk modulus of low-density cellular networks}%
%

\author{Marc Durand}%
%

\email{mdurand@ccr.jussieu.fr}%
%

\affiliation
{Mati\`{e}re et Syst\`{e}mes Complexes\\
UMR 7057 CNRS \& Universit\'{e} Paris 7 - Denis Diderot\\
Tour 33/34 - 2\`{e}me \'{e}tage - case 7056\\
2 Place Jussieu - 75251 Paris Cedex 05, France}%
%

\keywords{one two three}%
%

\pacs{89.75.Fb, 89.75.Hc, 82.70.Rr, 81.05.Zx}%
%

\begin{abstract}
We present an alternative derivation of upper-bounds for the bulk modulus of
both two-dimensional and three-dimensional cellular materials. For
two-dimensional materials, we recover exactly the expression of the
Hashin-Shtrikman (HS) upper-bound in the low-density limit, while for
three-dimensional materials we even improve the HS bound. Furthermore, we
establish necessary and sufficient conditions on the cellular structure for
maximizing the bulk modulus, for a given solid volume fraction. These
conditions are found to be exactly those under which the electrical (or
thermal) conductivity of the material reaches its maximal value as well. These
results provide a set of straightforward criteria allowing to address the
design of optimized cellular materials, and shed light on recent studies of
structures with both maximal bulk modulus and maximal conductivity. Finally,
we discuss the compatibility of the criteria presented here with the
geometrical constraints caused by minimization of surface energy in a real
foam.%
\end{abstract}%
%

\volumeyear{year}%
%

\volumenumber{number}%
%

\issuenumber{number}%
%

\eid{identifier}%
%

\received[Received text]{ 13/12/04}%
%

\revised[Revised text]{date}%
%

\accepted[Accepted text]{date}%
%

\published[Published text]{date}%
%

\startpage{101}%
%

\endpage{102}%
%

\maketitle

Cellular solids appear widely in nature and are manufactured on a large scale
by man. Examples include wood, cancellous bone, cork, foams for insulation and
packaging, or sandwich panels in aircraft. Material density, or solid volume
fraction, $\phi$, is a predominant parameter for the mechanical properties of
cellular materials. Various theoretical studies on the mechanical properties
of both two-dimensional (2D) and three-dimensional (3D) structures have been
attempted \cite{Weaire}. Unfortunately, exact calculations can be achieved for
cellular materials with simple geometry only \cite{Gibson}, and numerical
simulations \cite{Gibson}\cite{Roberts} or semi-empirical models
\cite{Christensen}\cite{Kraynik}\cite{Kraynik2} are required in order to study
the mechanical properties of more complex structures. However, expression of
bounds on the effective moduli can be established. Perhaps the most famous
bounds are those given by Z. Hashin and S. Shtrikman for isotropic heterogeneous
media \cite{Hashin-Shtrikman2D}\cite{Hashin-Shtrikman}. In particular, the
Hashin-Shtrikman bounds for the effective bulk modulus in the low-density
asymptotic limit ($\phi\ll1$) read:
\begin{equation}
0\leq\kappa^{\left(  2D\right)  }\leq\frac{E\phi}{4}\label{HS 2D}%
\end{equation}
for 2D cellular structures \cite{Hashin-Shtrikman2D}\cite{Torquato1}, and:%
\begin{equation}
0\leq\kappa^{\left(  3D\right)  }\leq\frac{4E\phi}{9}\frac{G+3K}%
{4G+3K}\label{HS 3D}%
\end{equation}
for 3D structures \cite{Hashin-Shtrikman}. $\kappa^{\left(  2D\right)  }$ and
$\kappa^{\left(  3D\right)  }$ are the actual bulk modulus respectively for 2D
and 3D structures, and $E,G,K$ are the Young modulus, shear modulus, and bulk
modulus of the solid phase, respectively. These three elastic moduli are
related by: $E=\frac{4KG}{K+G}$ for 2D bodies and by: $E=\frac{9KG}{3K+G}$ for
3D bodies.

The search for optimal structures maximizing some specific modulus (for a
given value of solid volume fraction $\phi$), is of evident practical
importance. In a recent study, Torquato et al. \cite{Torquato1}%
\cite{Torquato2}\ identified values of conductivity and elastic moduli of the
two-dimensional square, hexagonal, kagom\'{e} and triangular cellular
structures, and observed that the bulk modulus of these structures is equal to
the HS upper-bound value. The authors did not attempt to explain this result, although they noticed that such structures
under uniform compression deform without bend (affine compression). Are these structures the only structures with
maximal bulk modulus ? And if they are not, can we provide criteria on the structure of
"optimized" cellular materials ? More intriguingly, Torquato et al. noticed
that these structures present maximal electrical (or thermal) conductivity as
well. Is this feature caused by the particular symmetry of
the studied strutures, or is there an underlying relation between the conductivity and the bulk modulus of cellular materials ?
We shall answer to all these questions in the present study. Indeed, Durand \& Weaire
\cite{Durand1}\cite{Durand2} already established necessary and sufficient
conditions on the structure of cellular networks having maximal average conductivity.
A quite similar approach is used in this paper to show that the very same
conditions are also necessary and sufficient to maximize the bulk modulus of
an open-cell material. There are some evident similarities between the
constitutive laws (Ohm's law and Kirchhoff's laws) of electrical current in
wires and those of the thin beam theory, but complexity is increased in the
latter case, the scalar quantities $I$ (the electrical current) and $V$ (the
electrical potential) being replaced by the vectorial quantities $\mathbf{F}$
(force acting on a beam) and $\mathbf{u}$ (the displacement field).

\begin{figure}
[h]
\begin{center}
\includegraphics[
height=2.1718in,
width=2.3703in
]%
{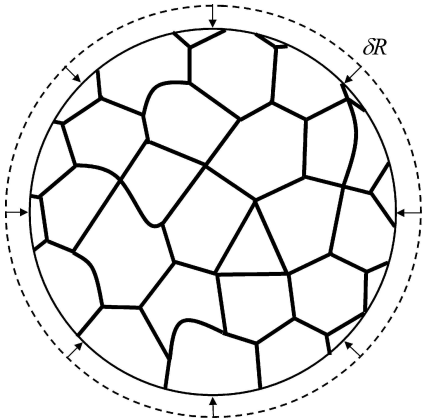}%
\caption{Circular portion of a 2D cellular network subjected to a uniform
radially oriented displacement $-\delta R\mathbf{e}_{r}$ of its boundary. The
network is made of thin beams with \textit{a priori} natural curvatures and
non-uniform cross-sections.\ We associate a bulk modulus for such a strain,
defined as: $\kappa^{(2D)}=\frac{R}{2}\frac{\delta P}{\delta R}$, where
$\delta P$ is the average load applied on the boundary.}%
\label{circular network}%
\end{center}
\end{figure}

We consider first the case of a 2D cellular material. We suppose its solid volume fraction
$\phi$ is sufficiently low, so the cell edges can be approximated as thin
beams. Beams \textit{a priori} can be naturally curved
and have non-uniform cross-sections, as long as the cross-sectional area
$s_{ij}\left(  l\right)  $ of each beam $\left(  i,j\right)  $ (where $l$
refers to the curvilinear coordinate along the beam, and $i$ and $j$ denote
the two nodes linked by the beam) is small compared with its length $l_{ij}$
squared. Let us isolate a circular portion of this material, of radius $R$,
and impose a uniform radial displacement $-\delta R\mathbf{e}_{r}$ on its
boundary ($\mathbf{e}_{r}$ is the radially oriented unit vector; the body is
under uniform tension when $\delta R<0$, and under uniform compression when
$\delta R>0$, see Fig. \ref{circular network}). We define the 2D bulk modulus
$\kappa^{\left(  2D\right)  }$ of this structure as:%
\begin{equation}
\frac{1}{\kappa^{(2D)}}=\frac{1}{\pi R^{2}}\frac{2\pi R\delta R}{\delta
P}=\frac{2}{R}\frac{\delta R}{\delta P},
\end{equation}
where $\delta P$\ is the average load applied on the boundary. We must point
out that the definition above is different from the usual definition of bulk
modulus: in the usual
definition, a uniform radial \textit{load} is applied on the surface of the material, while in the present definition a 
uniform \textit{displacement}
 is imposed on its surface (allowing to extend the notion of bulk-modulus to non-isotropic materials). However, the two definitions are
identical for 2D materials with square or hexagonal symmetry and for 3D
materials with cubic or isotropic symmetry.%

The expression of an upper bound can be easily established using the principle
of minimum potential energy: \textit{among all kinematically
admissible displacement fields (i.e. any displacement field twice continuously
differentiable satisfying the displacement constraints on the boundary), the
actual displacement (i.e. the one satisfying the equations of mechanical
equilibrium) is the one that makes the potential energy an absolute minimum}.
Let $\mathbf{u}^{\ast}\left(  \mathbf{r}\right)  $ be the displacement field
which satisfies the equations of equilibrium throughout the body and the
conditions on the boundary, $\mathbf{u}\left(  \mathbf{r}\right)  $ any
kinematically admissible displacement field, and $U\left(  \left\{
\mathbf{u}^{\ast}\left(  \mathbf{r}\right)  \right\}  \right)  $ and $U\left(
\left\{  \mathbf{u}\left(  \mathbf{r}\right)  \right\}  \right)  $ the
respective potential energy associated with these two displacement fields. Then, according to the principle of minimum potential energy:%
\begin{equation}
U\left(  \left\{  \mathbf{u}^{\ast}\left(  \mathbf{r}\right)  \right\}
\right)  \leq U\left(  \left\{  \mathbf{u}\left(  \mathbf{r}\right)  \right\}
\right)  . \label{inequality}%
\end{equation}
Let us choose as kinematically admissible displacement field:
$\mathbf{u}\left(  \mathbf{r}\right)  =-\frac{\delta R}{R}\mathbf{r}$, and let us
evaluate the potential energy associated with. We assume the
cross-section of each beam is sufficiently small so $\mathbf{u}\left(
\mathbf{r}\right)  $ is uniform on it (or equivalently, we suppose $\mathbf{u}\left(
\mathbf{r}\right)  $ is a macroscopic field which has a uniform value on the
beam cross-section). Thus, the stress tensor expressed in the local orthogonal
coordinate system has only one non-zero component: the axial-axial component.
Consider an infinitesimal piece of a given beam $\left(  i,j\right)  $, of
length $dl$ and cross-sectional area $s_{ij}(l)$. We denote $\mathbf{r}_{M}$ and
$\mathbf{r}_{M}\mathbf{+}d\mathbf{r}$ the position of its two ends. Their
relative displacement $\left(  \mathbf{u}\left(  \mathbf{r}_{M}\mathbf{+}%
d\mathbf{r}\right)  -\mathbf{u}\left(  \mathbf{r}_{M}\right)  \right)  $ is
colinear to the local tangent unit vector $\mathbf{t}_{ij}=\frac{d\mathbf{r}%
}{dl}$, meaning that the piece of beam deforms by axial compression only. The
force $\mathbf{F}_{ij}(l)$ acting on the surface $s_{ij}(l)$ is parallel to
$\mathbf{t}_{ij}$ and given by: $\mathbf{F}_{ij}(l)=Es_{ij}(l)\frac{\delta
R}{R}\mathbf{t}_{ij}$, where $E$ is the Young modulus of the solid phase. The
strain energy associated with such a deformation is $\frac{E}{2}%
s_{ij}(l)\left(  \frac{\delta R}{R}\right)  ^{2}$. Invoking additivity of the
potential energy :%
\begin{equation}
U\left(  \left\{  \mathbf{u}\left(  \mathbf{r}\right)  \right\}  \right)  =%
{\displaystyle\sum\limits_{\left(  i,j\right)  }}
{\displaystyle\int\limits_{0}^{l_{ij}}}
\frac{E}{2}s_{ij}(l)\left(  \frac{\delta R}{R}\right)  ^{2}dl
\end{equation}
(where the discrete sum is carried out on all the beams $\left(  i,j\right)
$), and introducing the volume fraction of solid:%
\begin{equation}
\phi=%
{\displaystyle\sum\limits_{\left(  i,j\right)  }}
{\displaystyle\int\limits_{0}^{l_{ij}}}
s_{ij}(l)dl/\pi R^{2}, \label{volume fraction}%
\end{equation}
we obtain: 
\begin{equation}
U\left(  \left\{  \mathbf{u}\left(  \mathbf{r}\right)  \right\}
\right)  =\frac{\pi}{2}E\phi\left(  \delta R\right)  ^{2}. \label{trial energy}
\end{equation}
On the other hand, the actual potential energy $U\left(  \left\{  \mathbf{u}^{\ast}\left(
\mathbf{r}\right)  \right\}  \right)  $ is equal to half the work done by the
external forces \cite{Love}:%
\begin{equation}
U\left(  \left\{  \mathbf{u}^{\ast}\left(  \mathbf{r}\right)  \right\}
\right)  =\frac{1}{2}\delta P2\pi R\delta R=2\pi\kappa^{\left(  2D\right)
}\left(  \delta R\right)  ^{2}. \label{actual energy}
\end{equation}
Comparison of Eqs. \ref{trial energy} and \ref{actual energy} finally leads to an upper-bound for the bulk modulus:%
\begin{equation}
\kappa^{\left(  2D\right)  }\leq\frac{E\phi}{4}. \label{upper bound 2D}%
\end{equation}
The same argumentation can be used for 3D
open-cell structures: in that case, we find that the bulk modulus $\kappa^{(3D)}$ associated
with a spherical portion of material of radius $R$, and defined as:%
\begin{equation}
\frac{1}{\kappa^{(3D)}}=\frac{1}{\frac{4}{3}\pi R^{3}}\frac{4\pi R^{2}\delta
R}{\delta P}=\frac{3}{R}\frac{\delta R}{\delta P},
\end{equation}
is bounded as it follows:
\begin{equation}
\kappa^{\left(  3D\right)  }\leq\frac{E\phi}{9}.
\end{equation}
The upper-bound value we obtain is then lower than the HS upper-bound
value \ref{HS 3D}, giving a sharpest estimation of the actual bulk modulus value.

\subsection{Criteria for maximal bulk modulus}

The upper bounds established above are optimal bounds, i.e. there 
exist cellular materials with maximal bulk modulus value. Criteria on the
structure of such materials can be provided. Indeed, we show in the
following that the bulk modulus equals the upper-bound value \textit{if and
only if} the three following conditions are simultaneously satisfied:

\begin{enumerate}
\item[a)] All the edges are straight.

\item[b)] Each edge has a uniform cross-section area: $s_{ij}(l)=s_{ij}$.

\item[c)] Every junction $(i)$ between edges satisfies $\sum_{j}%
s_{ij}\mathbf{e}_{ij}=\mathbf{0}$, where $\mathbf{e}_{ij}$ are
outward-pointing unit vectors in the directions of adjoining edges.
\end{enumerate}

\begin{figure}
[h]
\begin{center}
\includegraphics[
height=1.0677in,
width=1.858in
]%
{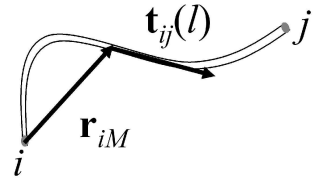}%
\caption{Schematic of a paticular beam $\left(  i,j\right)  $. $l$ denotes the
curvilinear coordinate of a given point $M$ along the beam. $\mathbf{r}_{iM}$
and $\mathbf{t}_{ij}$ are respectively the position vector of point $M$ taken
from node $i$ and the local tangent unit vector.}%
\label{curved beam}%
\end{center}
\end{figure}

The demonstration is straightforward: according to the principle of minimal
potential energy and the uniqueness of the actual displacement field, the
inequality \ref{inequality} becomes a strict equality if and only if the trial
displacement field $\mathbf{u}\left(  \mathbf{r}\right)  =-\frac{\delta R}%
{R}\mathbf{r}$ is the actual displacement field satisfying the equations of
mechanical equilibrium. Inspection of force and moment balances along each
beam and at each junction leads to the three necessary and sufficient
conditions stated above. Let us make this precise. Consider a specific beam
$\left(  i,j\right)  $ (see Fig. \ref{curved beam}); at equilibrium, the
moments of forces acting on it must balance. Choosing as referencing point for the moments the node
$i$, and denoting $\mathbf{r}_{iM}=$ $\mathbf{r}_{M}-\mathbf{r}_{i}$, where
$\mathbf{r}_{i}$ and $\mathbf{r}_{M}$ are the respective position vectors of
node $i$ and of any point $M$ belonging to the beam, we obtain: $\mathbf{r}%
_{iM}\times\mathbf{F}_{ij}(l)=Es_{ij}(l)\frac{\delta R}{R}\mathbf{r}%
_{iM}\times\mathbf{t}_{ij}=\mathbf{0}$. Thus, the tangent unit vector
$\mathbf{t}_{ij}$ must be parallel to the position vector $\mathbf{r}_{iM}$
for any point $M$ belonging to the beam, leading to condition a). The forces
acting on any piece $\triangle l$ of the straight beam $\left(  i,j\right)  $
must balance as well: $E\frac{\delta R}{R}s_{ij}(l)=E\frac{\delta R}{R}%
s_{ij}(l+\triangle l)$ what immediately leads to condition b). Finally,
mechanical equilibrium at every junction $i$ is satisfied if: $\sum
_{j}\mathbf{F}_{ij}=\mathbf{0}$, with $\mathbf{F}_{ij}=-Es_{ij}\frac{\delta
R}{R}\mathbf{e}_{ij}$, leading to condition c). The moment balance at every
junction is automatically satisfied when conditions a), b), c) are fulfilled,
since the force acting on each straight beam is then axially oriented.%
Furthermore, we check that the geometrical constraint on angles between adjoining edges is also satisfied; if $\mathbf{e}_{ij}^{\prime}$ denotes the unit
vector parallel to the beam $\left(  i,j\right)  $ \textit{after} deformation
and $\mathbf{r}_{ij}=\mathbf{r}_{j}-\mathbf{r}_{i}$, then:%
\begin{equation}
\mathbf{e}_{ij}^{\prime}=\frac{\mathbf{r}_{ij}+\mathbf{u}\left(
\mathbf{r}_{j}\right)  -\mathbf{u}\left(  \mathbf{r}_{i}\right)  }{\left\Vert
\mathbf{r}_{ij}+\mathbf{u}\left(  \mathbf{r}_{j}\right)  -\mathbf{u}\left(
\mathbf{r}_{i}\right)  \right\Vert }=\frac{\left(  1-\frac{\delta R}%
{R}\right)  \mathbf{r}_{ij}}{\left(  1-\frac{\delta R}{R}\right)  \left\Vert
\mathbf{r}_{ij}\right\Vert }=\mathbf{e}_{ij},
\end{equation}
what proves the material deforms by affine compression, and the angles between
edges are preserved.

\subsection{Some comments}

We first summarize the limits of the theory: the solid volume fraction is supposed to be low
enough for the thin beam theory to be valid. Moreover the relative imposed
displacement $\frac{\delta R}{R}$ must be small enough so that
Hooke's law can be used and no mechanical Euler instability occurs when the
body is under compression.

Conditions a), b), c) are the necessary and sufficient
conditions to maximize the average conductivity of a network of thin wires as
well \cite{Durand2}. Why structures satisfying these conditions have both
maximal bulk modulus and maximal conductivity ? When the three conditions are fulfilled, the force acting
to each beam is then parallel to it, and the corresponding deformation of each beam
is an axial compression (or tension); no bending or twisting occurs. The
"flow" of stress is parallel to the beams, as for the electric courant, and
the geometry defined by the three conditions corresponds to the most homogeneous
distribution of constraints and currents through the whole structure.

\begin{figure}
[h]
\begin{center}
\includegraphics[
height=1.4587in,
width=3.0735in
]%
{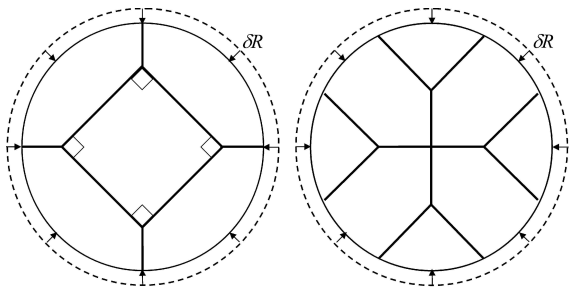}%
\caption{Examples of cellular networks for wich condition c) is not satisfied
and still deform by affine compression (no bending or twisting of the beams).
However their respective bulk modulus is strictly lower than the upper bound
value $\frac{E\phi}{4}$.}%
\label{examples}%
\end{center}
\end{figure}

We must point out that condition c) is sufficient for having no bending in a
structure for which conditions a) and b) are fulfilled, but not necessary: 
there do exist structures which do not satisfy condition c) and 
which deform under compression without bending (e.g. see structures of Fig. \ref{examples}). Nevertheless, the
bulk modulus of such structures will be below the upper-bound value; condition c)
must be satisfied in order to have maximal bulk modulus.%

Furthermore, it is worth noticing that the three conditions are independent of
the connectivity of the junctions. As a consequence, there is an infinity of
structures with maximal bulk modulus. Indeed, various examples of structures
with maximal bulk modulus can be found in literature: we can cite the square,
hexagonal, triangular, kagom\'{e} networks \cite{Torquato1}\cite{Torquato2} as
2D structures and the cubic \cite{Gibson} and Kelvin networks
\cite{Christensen}\cite{Kraynik}\cite{Zhu} as 3D structures. All these
structures satisfy the necessary and sufficient conditions a), b), c), in
agreement with the work presented here. Numerical simulations of random 2D
\cite{Torquato1}\cite{Silva} and 3D \cite{Kraynik3} isotropic cellular
materials have been also generated. As expected, the bulk modulus of those
materials is found to be always lower than the respective upper bound values.

As concluding remarks, let us discuss the consequences of the criteria
presented here on the mechanical properties of real foams. Foam in the
low-density limit is a particular cellular material: usually its preparation
involves a continuous liquid phase that eventually solidifies. Therefore its
structure is controlled by minimization of surface energy, leading to
geometrical rules known as the Plateau's laws \cite{Weaire}, which can be
summarized as it follows:

- edges in a 2D foam meet in threefold junctions with equal angles of 120$%
{{}^\circ}%
$.

\begin{figure}
[h]
\begin{center}
\includegraphics[
height=1.5807in,
width=1.5276in
]%
{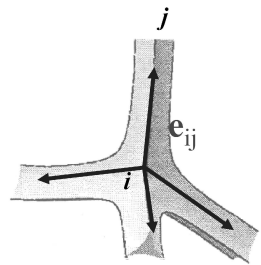}%
\caption{In a real 3D foam, the edges (also called Plateau borders) meet in
fourfold junction with equal angles, corresponding to the tetrahedral angle:
$\arccos\left(  -\frac{1}{3}\right)  \simeq109,5{{}^\circ}$.}%
\label{Plateau}%
\end{center}
\end{figure}

- lamellae in a 3D foam meet in threefold lamella junctions (usually called
Plateau borders) with equal angles of 120$%
{{}^\circ}%
$, and Plateau borders meet in fourfold junctions with the tetrahedral angle:
$\arccos\left(  -\frac{1}{3}\right)  \simeq109,5%
{{}^\circ}%
$ (see Fig. \ref{Plateau}).%

As a consequence, condition c) is always satisfied in a real foam. Usually,
condition b) is nearly satisfied as well. The validity of condition a) is more
delicate: while it is still possible to build 2D foams satisfying
simultaneously conditions a), b), c) and the Plateau's laws (e.g. the
hexagonal honeycomb), this is no longer true for the 3D case: no cell in a 3D
foam is a simple polyhedron with straight edges, because a planar polygon
cannot have all angles equal to the tetrahedral angle. Consequently, edges in
a real foam must be curved, violating condition a), and the bulk modulus and
the average conductivity of a solid open-cell foam (i.e. a foam where the
lamellae broke up during solidification) is always strictly lower that the
corresponding bounds. While the conductivity drop is not really significant
when the edges are slightly curved \cite{Durand1}, the bulk
modulus value can be dramatically decreased, because beams can easily bend or twist. We
conclude that 3D cellular materials manufactured by some foaming process are
probably not the most relevant for the design of high-bulk modulus/low-density structures.

J. Robiche, J.-F. Sadoc and D. Weaire are deeply acknowledged for useful
discussions and inspection of this work.

\end{document}